\documentclass[english,preprintnumbers,amsmath,amssymb]{revtex4}

\usepackage[T1]{fontenc}
\usepackage[latin9]{inputenc}
\usepackage{amsmath}
\usepackage{graphicx}
\usepackage{amsthm,amsfonts,amssymb,times,bbm}
\usepackage[center]{subfigure}



\begin{document}

\title{Discrete Wigner Function Reconstruction and Compressed Sensing}

\author{Jia-Ning Zhang}
\email{galileo@mail.nankai.edu.cn}
\author{Lei Fang}
\author{Mo-Lin Ge}
\affiliation{Theoretical physics division, Chern Institute of Mathematics, Nankai University,
Tianjin 300071, China}

\date{\today}

\begin{abstract}
A new reconstruction method for Wigner function is reported for quantum
tomography based on compressed sensing. By analogy with computed
tomography, Wigner functions for some quantum
states can be reconstructed with less measurements utilizing
this compressed sensing based method.
\end{abstract}

 \maketitle

Wigner function(WF), a quasi-probability distribution in phase
space,  was first introduced to describe quantum state in quantum
mechanics by E. P. Wigner\cite{wigner}. And later, it was extended
to classical optics and signal processing. Since its birth, a great number of applications
 have been conducted in different fields. As for the original quantum case, like a continuous one-dimensional
quantum system, the Wigner function is defined as
\begin{eqnarray}
W(x,p)\equiv\frac{1}{\pi}\int_{-\infty}^{\infty}d\xi
\exp(-2ip\xi)\langle x+\xi|\hat{\rho}|x-\xi\rangle
\end{eqnarray}
where $x$ and $\xi$ are positions and $p$ is momentum, $\hat{\rho}$ denotes
the density operator, and $\hbar$ is set to 1 for simplicity.

 One of the advantage for the Wigner formalism is that a tomographic
scheme
which could be used to reconstruct the WF. On the other hand, in medical tomography, like X-ray
computed tomography, X-ray photons transmitted through the patient
along projection lines from one side, and detectors measured the
number of transmitted photons on the other side, then from the
distribution of X-ray photons in different incident angles, linear
attenuation coefficients in the slice being imaged are recovered. Likewise, in phase space tomography, the WF tomographic reconstruction is
more or less the same, based on a set of intensity
measurements\cite{vogelpra}. This is because both
of the two tomographic schemes share the same mathematical basis ---
Radon transform(Eqn.(2)) and its inverse(Eqn.(3)).

\begin{eqnarray}
pr(x_{\phi},\phi)= \int\int f(x,y)\delta(x_{\phi}-x\cos\phi -
y\sin\phi)dx dy
\end{eqnarray}

\begin{eqnarray}
f(x,y)=-\frac{1
P}{2\pi^{2}}\int_{0}^{\pi}\int_{-\infty}^{+\infty}\frac{pr(x_{\phi},\phi)dx_{\phi}d\phi}{(x\cos\phi
+y\sin\phi-x_{\phi})^{2}}
\end{eqnarray}
 Where $P$ denotes the Cauchy's principle value, $pr(x_{\phi},\phi)$ denotes the measured distribution, in $f(x,y)$ is the quantity in tomography to be recovered. In tomography, objective function is reconstructed using
back-projection algorithm for the inverse Radon
transformation(Eqn.(3)). However, such a algorithm has some
drawbacks, for example, many measurements are needed to achieve a
relatively good recovery. To reduce the dose of radiation on the
patient in CT or make a more efficient estimate of quantum state in
phase space tomography, the amount of measurements should be as
small as possible but guarantee a good enough recovery.
Nevertheless, according to the well-known Nyquist-Shannon sampling
theorem\cite{unser}, it seems impossible to depress the measurements
without giving a negative impact on the resolution.

Recently, with the arise of a novel theory called compressed sensing
(compressive sampling or CS) \cite{candes}, people find a
fundamentally new approach to acquire data.
 It demonstrates that if the images or signals are sparse, we can recover them from what was
  believed to be highly incomplete measurements.
 In fact, it is just the application to medical tomography that started the seminal paper in CS.
For example, in CT, X-ray images of human body are sparse in the
wavelet and
  gradient representation, then compressed sensing is applied to reduce the irradiation and obtain clearer medical images\cite{sidky,Chen}.
And then an immediate question would be is it possible to reconstruct WF
in phase space tomography via compressed sensing?
In fact, there exists infinite family of quasi-probability distribution in phase, including Husimi-Kando Q-Function, Glauber-Sudarshan P-Function and so on,
but WF is chose for it is directly related to the quadrature histograms and can be easily measured through experiments.

 In this brief report,  we explore the similarity between phase space tomography and computed tomography, and report a
 new discrete WF
reconstruction method based on compressed sensing. We mention that
some compressed sensing protocols have been proposed to reconstruct
density matrix though quantum
tomography\cite{Grossprl}.

Suppose that a signal $x\in \mathbb{C}^{N}$ is sparse in some basis
 $x=\Psi x'$, which means $x'$ has $k(k\ll N)$ nonzero
entries. Then a measurement matrix $\Phi\in \mathbb{C}^{m\times N}$
is used to sense $x$ and obtain a measurement vector $y\in
\mathbb{C}^{m}$.
 The central idea of compressed
sensing is that it is possible to reconstruct sparse signals of
scientific interest accurately and sometimes exactly by a number of
incoherent sampling which is far smaller than N. In
general, such a recovery of $x$ by performing
$\ell_{0}$-minimization is a NP-hard problem. An alternative
procedure called $\ell_{1}$-minimization is usually used which says

\begin{equation}
\min_{x'}\|x'\|_{\ell_{1}}\qquad\textrm{subject to}\qquad
y=\Phi\Psi x'=\Upsilon x',
\end{equation}
This is a convex optimization problem, and many numerical
algorithms apply for its solution.

Before showing the reconstruction method, we note that there are two
main differences between conventional medical tomography and the
quantum tomography. First, WF is a pseudo probability distribution,
may have negative values, while the attenuation coefficients in CT
images are always nonnegative. Second, for a given angle, a complete
probability density for the quadrature operator is measured due to
the quantum mechanical property, but in medical tomography,
detectors' number can be adjusted. On the other hand, similar to the
medical case, at the heart of a possible CS application to phase
space tomography is to find a suitable sparse representation basis.
In real computed tomography, images are generally extended
distributions which violate the prerequisite of $\ell_{1}$-based
algorithms. But other sparseness property (gradient sparse) and
reconstruction algorithm(TV-algorithm)\cite{sidky} are found to
 reconstruct certain tomographic images which are relatively
 constant over extended areas.  While due to its configuration's flexibility
, it is hard even impossible to find a unitive sparse basis for WF
in general. However, for some special applications, WF might share
the same sparse basis. And in contrary to the medical images, some
quantum states have finite values on a small region in phase space
with zero or nearly zero values elsewhere, so have been sparse in
this pseudo-pixel representation. Other situations may be completely
different, for WF is not sparse in phase space representation. A basis
transform like Fourier transform, wavelet transform may bring WF to
a sparse representation.

Although it is also possible to reconstruct continuous Wigner function, but discretization must be taken before it can be tackled in practise.
So in this report we adopt a discrete quantum tomography
model. But we note that it should be straightforward to transport our idea to an continuous one. And indeed such an attempt has been taken in information science\cite{Chen}. The quantum state
tomographic model for discrete Wigner functions
first proposed by U. Leonhardt\cite{leonhardt}. We note that continuous and discrete
quantum tomography are the same in principle. In finite quantum
systems, position and momentum take discrete values in
$\mathbb{Z}_{d}$. In Leonhardt's discrete quantum tomography picture,
a WF in a $d$-dimensional(odd) Hilbert space was proposed(Eqn.(6)) and
phase was quantized to take place of momentum in continuous variable
quantum systems.

\begin{eqnarray}
W(m,\mu)=\frac{1}{d}\sum_{n}\exp(-\frac{4\pi i}{d}\mu n )\langle
m-n|\hat{\rho}|m+n \rangle
\end{eqnarray}
Here, $m, n$ denotes the spins or particle numbers, $\mu, \nu$
denotes the quantized phases. Similar to the continuous case, the
observable distribution is the overlap of measured state's WF and
the observable state's WF
 in phase space.
\begin{eqnarray}
&&pr(\mu_{0}, \tau)=\sum_{\mu, m}
W(m,\mu)\delta(\mu-\mu_{0}+2m\tau;d)\nonumber \\
&&pr(m_{0}, \tau)=\sum_{\mu, m} W(m,\mu)\delta(m-m_{0};d)\delta(m-m_{0};a).
\end{eqnarray}
Here, $m_{0},\mu_{0}\in\mathbb{Z}_{d}, pr(\mu_{0}, \tau)$ denotes
the probability distribution, $W(m,\mu)$ denotes the WF to be
recovered, $\tau\in\mathbb{Z}_{d}$, $m, m_{0}$ represent the spins
or particle numbers, $\mu, \mu_{0}$ represent the quantum phases and
$\delta(k;a)$ is the modular Kronecker symbol(equals 1 when
$k$\textrm{=0 (mod a}), and zero otherwise)

In our simulation, a WF for finite dimensional coherent state
$||\alpha|\textrm{e}^{i\phi}\rangle_{s}$ (s=d-1)\cite{Miranowicz} is
used as an example to illustrate the CS based reconstruction
algorithm.

\begin{eqnarray}
W(m,\mu)&=&\sum_{k=M+1}^{s}\nonumber\\
&\times&\frac{\exp[i(2k-M-s-1)(2\pi\mu/d-\phi+\pi/2)]}{[k!(M-k-s-1)!]^{1/2}}G_{1k}\nonumber\\
&+&\sum_{k=0}^{M}\frac{\exp[i(2k-M)(2\pi\mu/d-\phi+\pi/2)]}{[k!(M-k)!]^{1/2}}G_{0k}
\end{eqnarray}

where
\begin{eqnarray}
G_{\eta
k}&=&\frac{(s!)^{2}}{(s+1)^{3}}\sum_{p=0}^{s}\sum_{q=0}^{s}\nonumber\\
&\times&\exp[i(x_{q}-x_{p})|\alpha|]
\frac{\textrm{He}_{k}(x_{p})\textrm{He}_{M-k+\eta(s+1)}(x_{q})}{[\textrm{He}_{s}(x_{p})\textrm{He}_{s}(x_{q})]^{2}}\nonumber
\end{eqnarray}
with $\eta=0, 1$, $M\equiv2n\textrm{mod}(d)$, $\textrm{He}_{s+1}(x)$
is the Hermite polynomial
($\textrm{He}_{n}(x)\equiv2^{-n/2}H_{n}(x/\sqrt{2})$) and $x_{k}$ is
the $k$th root of $\textrm{He}_{s+1}(x)=0$.

It should be pointed out that different from classical signal
processing (the sensing matrix elements can be independently
selected from a random distribution), only the rows of the sensing
matrix can be randomly measured in quantum tomography. The sensing
matrix was randomly selected from the rows of full measurement
matrix (modular Kronecker in Eqn.(7)). Then, we get the precise probability distribution by multiplying the sensing matrix and the WF.
 The problem considered now is to reconstruct a state
from measured insufficient data. Computer simulation has been
conducted to demonstrate the effectiveness of the aforementioned
scheme.  All the recovery processes were implemented in Java and C.
For $\ell_{1}$-minimization, we adopt a linearized bergman iteration
algorithm\cite{yin}.

To simulate the sampling process, we assign a number to each row of
the full measurement matrix, then computer generated a series of
random numbers representing the rows we select to form the sensing
matrix. For simplicity, the experiment error was totally ignored.
 So in this pseudo pixel
representation, WF can be just roughly reconstructed. This is partly due to measurement process itself, for only a small
fraction of the WF is sensed once, which means the sensing matrix
is not dense enough. To overcome this obstacle, two
approaches might help: finding another sparse representation or using a
 denser sensing matrix(measurement quantum state, for example,
coherent state). For example, for some type of mixed state ensemble,
WF varies along a particular direction when we rearrange the
two dimensional WF into a one dimensional object carefully, it is
sparse in a discrete cosine Fourier representation, which means this basis is its sparse basis.

\begin{figure}
\renewcommand{\figurename}{FIG.}
\centering \subfigure[]{
\label{fig:subfig:a}
\includegraphics[width=2in]{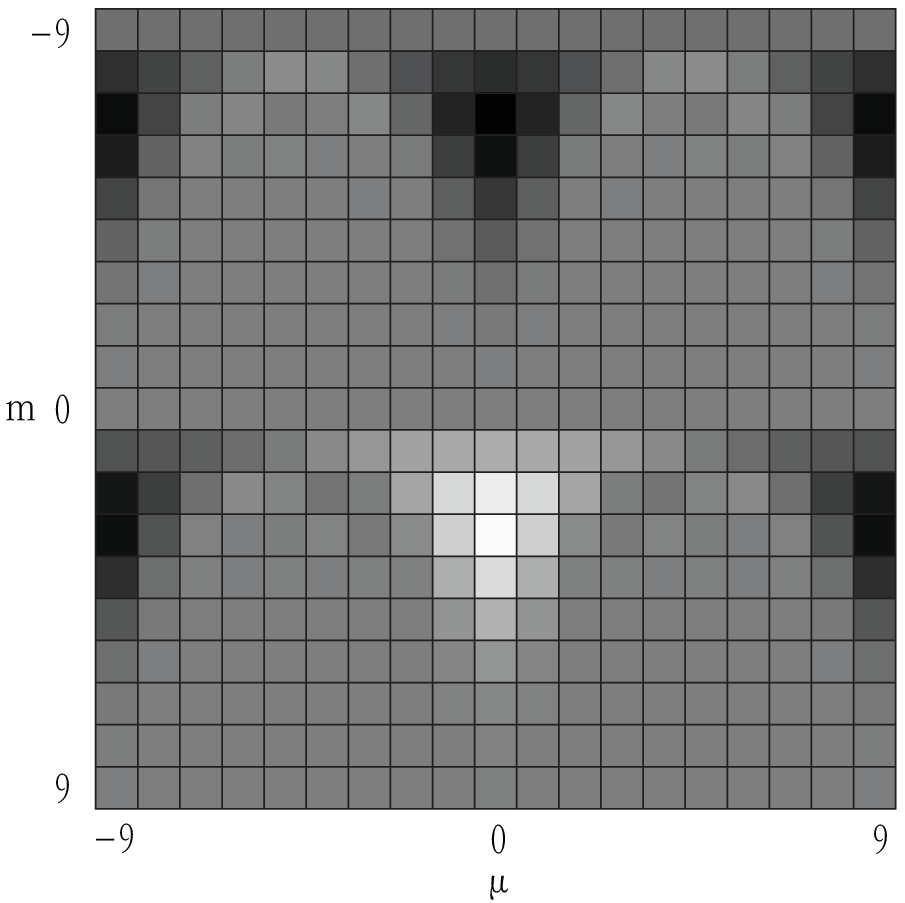}}
\hspace{1in}%
\subfigure[]{
\label{fig:subfig:b}
\includegraphics[width=2in]{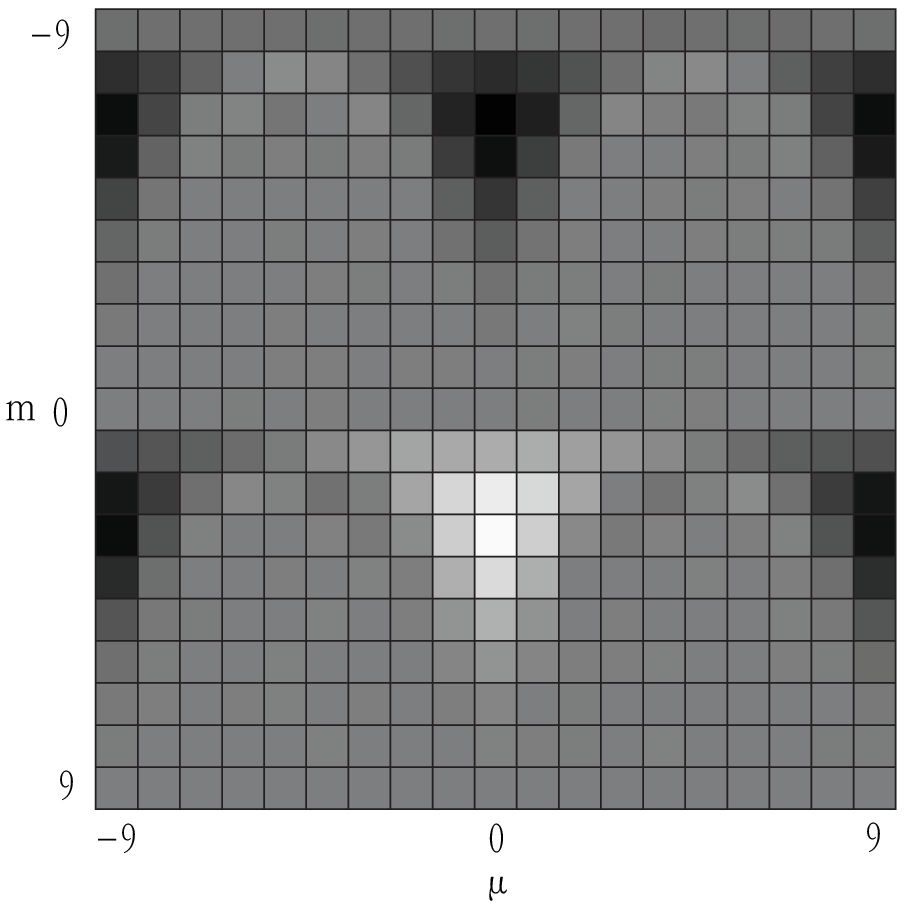}}
\caption{  Darker regions correspond to higher values, the regions between dark and white are dominated by zero or nearly zero values
.(a)Accurate WF for finite dimensional coherent
state(Eqn.(8), $d=19$, $|\alpha=1.472\rangle$). (b) Recovered WF via
compressed sensing, 285 rows have been taken from total
measurement matrix as the sensing
matrix.} \label{fig:subfig}
\end{figure}

 In spite of some drawbacks, when we deal with a high dimensional
quantum state reconstruction or what we need is just a rough
knowledge of WF, our reconstruction algorithm would show its
advantage, it makes the measurements more economical. And any other prior
knowledge could also be helpful to recovery process, for example,
the symmetry of the WF. By exploring the correspondence between
compressed sensing and tomography, we have shown the possibility of
a CS-based WF reconstruction method.  In addition, most convex
optimization algorithms related to CS theory deal with real signal.
And in our case, WF is real function. But when we need to reconstruct complex signals or images, for example, ambiguity function(AF)\cite{testorf}
which is complex in general, problems appear. Here, we mention that
G.Zweig proposed an algorithm\cite{zweig} processing complex
signals, which converted the the usual $\ell_{1}$-minimization in
complex space to a linear programming problem in real space.

In conclusion, we have described a reconstruction method based on
compressed sensing to recover the WF in phase space tomography. We
regard the Wigner function as an image in phase space. The values of
WF on different sites corresponds to pixels' gray values. Our
numerical simulation has shown good recovery for a sparse discrete
WF with usually supposed insufficient measurements. Such
a method can also be easily extended to reconstruct electromagnetic
field's WF with just some straightforward revisions. We hope that our
scheme can serve as a motivation for other quantum information or
phase space optics researches. Possible relations to the popular
maximum-likelihood reconstruction in quantum tomography still need
to be explored. By combination with other reconstruction ideas,
we believe a more efficient and precise phase recovery should be
possible.

This work is supported by the NSF of China(Grant No. 11075077) and also partly by the
SRFDP of China(Grant No. 200800550015).

\end{document}